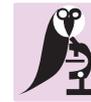

DIAGNOSTIC PATHOLOGY

SOFTWARE  Open Access

# Open source tools for management and archiving of digital microscopy data to allow integration with patient pathology and treatment information

Matloob Khushi[1*], Georgina Edwards[2], Diego Alonso de Marcos[2], Jane E Carpenter[1], J Dinny Graham[3] and Christine L Clarke[1,3]

## Abstract

**Background:** Virtual microscopy includes digitisation of histology slides and the use of computer technologies for complex investigation of diseases such as cancer. However, automated image analysis, or website publishing of such digital images, is hampered by their large file sizes.

**Results:** We have developed two Java based open source tools: Snapshot Creator and NDPI-Splitter. Snapshot Creator converts a portion of a large digital slide into a desired quality JPEG image. The image is linked to the patient's clinical and treatment information in a customised open source cancer data management software (Caisis) in use at the Australian Breast Cancer Tissue Bank (ABCTB) and then published on the ABCTB website (http://www.abctb.org.au) using Deep Zoom open source technology. Using the ABCTB online search engine, digital images can be searched by defining various criteria such as cancer type, or biomarkers expressed. NDPI-Splitter splits a large image file into smaller sections of TIFF images so that they can be easily analysed by image analysis software such as Metamorph or Matlab. NDPI-Splitter also has the capacity to filter out empty images.

**Conclusions:** Snapshot Creator and NDPI-Splitter are novel open source Java tools. They convert digital slides into files of smaller size for further processing. In conjunction with other open source tools such as Deep Zoom and Caisis, this suite of tools is used for the management and archiving of digital microscopy images, enabling digitised images to be explored and zoomed online. Our online image repository also has the capacity to be used as a teaching resource. These tools also enable large files to be sectioned for image analysis.

**Virtual Slides:** The virtual slide(s) for this article can be found here: http://www.diagnosticpathology.diagnomx.eu/vs/5330903258483934

**Keywords:** Snapshot Creator, NDPI-Splitter, Virtual microscopy, Digital slides, Caisis, Deep Zoom

## Background

Over the past decade there has been a marked increase in the use of virtual microscopy. Digital slides offer many benefits over traditional microscopy, such as ease of access, archiving, annotation and sharing. Automatic identification and percentage calculation of malignant/cancer regions of hundreds of archive slides have become possible by the use of data mining analysis tools [1-3]. Multiple digital slide images can be opened and analysed at the same time. For example Hematoxylin and eosin (H&E) and Periodic acid-Schiff (PAS) stained slides can be compared on the same screen, which is not possible in traditional microscopy. As unlimited users can examine specimens at the same time and this is independent of access time, many institutions have started teaching virtual microscopy as part of their regular histology course while others are considering moving in this direction [2,4,5].

* Correspondence: mkhushi@uni.sydney.edu.au
[1]Australian Breast Cancer Tissue Bank, University of Sydney at the Westmead Millennium Institute, Darcy Road, Westmead, NSW 2145, Australia
Full list of author information is available at the end of the article

The advantages of digitizing pathology slides are counterbalanced by the very large file sizes that are generated. A typical scanned slide at 400x magnification can be as large as 5 Giga bytes (0.25 μm/pixel) or even greater for higher resolutions. Such large file sizes hamper the downloading, viewing and analysis of digital slide images.

Proprietary image viewers such as Hamamatsu's NDP. view or Aperio's ImageScope only allow the user to take manual snapshots of the image being viewed, thereby limiting the maximum resolution to the resolution of the screen. For example, if the user's display resolution is set at 1680x1050, then the maximum resolution of a snapshot would be 17.6 Mega pixels. This is insufficient for snapshots to be published as zoomable slides on the website, which require snapshots at a resolution of 45 Mega pixels or more. Similarly there is no tool that is able to split the digital slides scanned by the Hamamatsu NanoZoomer into smaller sections. Therefore, this study had two objectives; firstly, to enable the publishing of snapshots of virtual slides on a tissue bank website and the building of a searchable digital microscopy database, and secondly, producing smaller images from large virtual slides to enable easy analysis and handling by analysis software such as Metamorph® (Molecular Devices, USA) or MATLAB (MathWorks, U.S.A.). In order to achieve these objectives, we have developed two open source tools: 'Snapshot Creator' and 'NDPI-Splitter'.

## Implementation

The tools are designed for digital images obtained on the Hamamatsu NanoZoomer Digital Pathology (NDP) System (Hamamatsu Photonics K.K. Japan), in their proprietary NDPI file format. NDPI-Splitter and Snapshot Creator are developed in Java using Standard Widget Toolkit (SWT), however it is Windows dependent because of the use of Hamamatsu SDK, available from Hamamatsu under their licensing agreement, for manipulating. NDPI files. It also uses JAI 1.1.3 available from http://ndpi-splitter.googlecode.com/files/jai-1_1_3-lib-windows-i586-jre.exe and JAI Image IO 1.1 available from http://ndpi-splitter.googlecode.com/files/jai_imageio-1_1-lib-windows-i586-jre.exe. Apache Ant (http://ant.apache.org/) is used to build the projects.

### Software architecture

Both NDPI Splitter and Snapshot Creator use the same underlying java classes to interact with the proprietary NDPI file format. These classes provide a Java wrapper around the Hamamatsu Software Development Kit (SDK) using Java Native Access (JNA) (Figure 1). The SDK provides low-level access to the image. The Java wrapper classes provide an easy to use Application Programming Interface (API) which includes operations to find out image dimensions and request portions of the image.

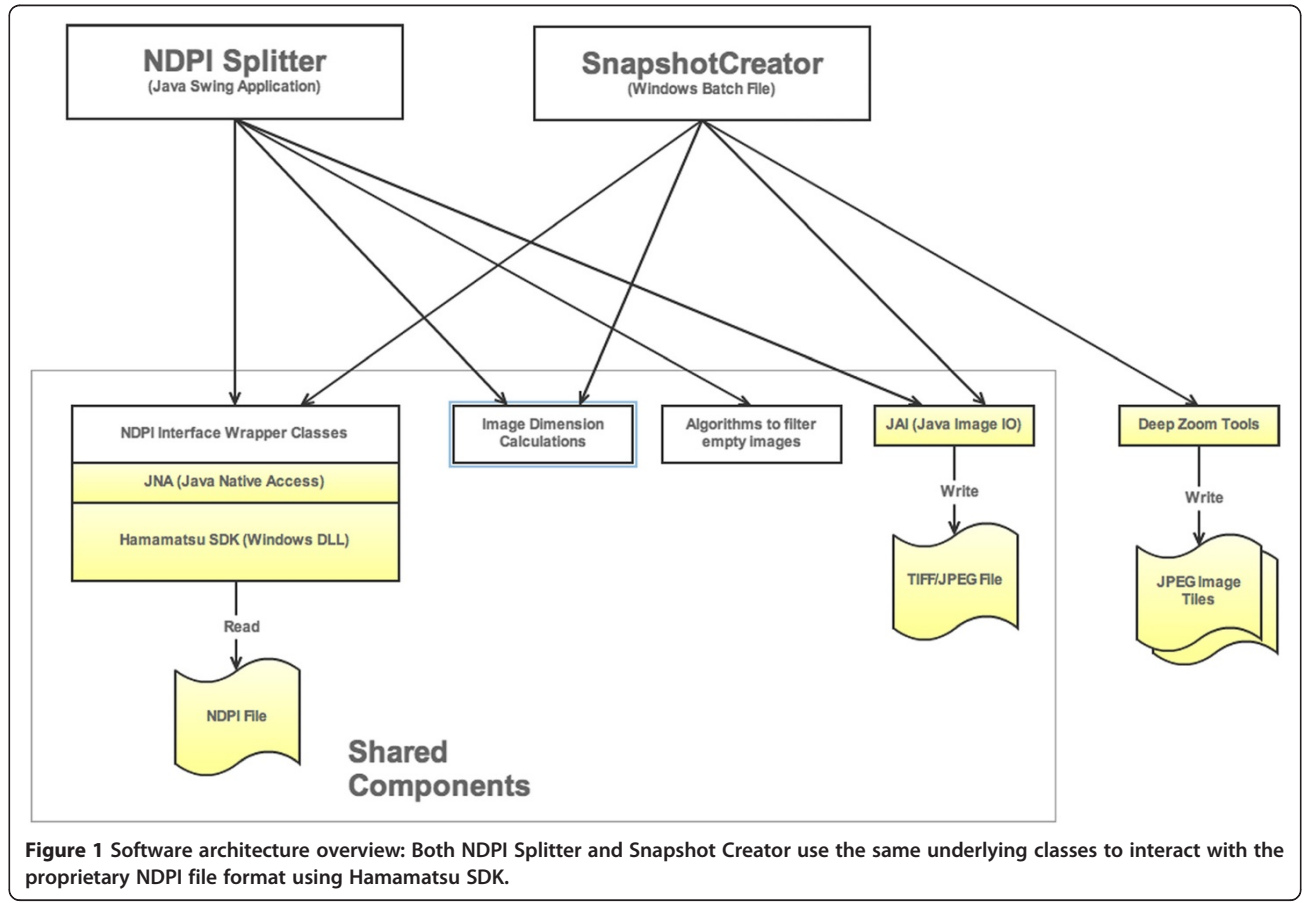

**Figure 1 Software architecture overview:** Both NDPI Splitter and Snapshot Creator use the same underlying classes to interact with the proprietary NDPI file format using Hamamatsu SDK.

NDPI Splitter is a Java Swing based graphical user interface (GUI) application. It uses the library classes described above to determine the size and dimensions of the image, then uses this information to calculate how to split up the image. It also includes a module to perform the filtering of "empty" images.

Snapshot Creator is a windows batch file with supporting Java classes. The Java classes use the library classes described above to determine size and dimensions of the image, and to extract the required size of the image. The batch file then uses the Deep Zoom converter tool to prepare the JPEG image for viewing in Deep Zoom.

### Deep Zoom

To facilitate panning and zooming of images on our website, we have used the Microsoft Deep Zoom [6] library. The JPEG converted images, produced by Snapshot Creator, are fed through the Deep Zoom converter tool to create tiles of images at various resolutions. Deep Zoomed images are then published on the website and linked to our customised version of the Caisis database [7].

Deep Zoom is acknowledged to be one of the most effective zooming technologies in current use [6]. Deep Zoom implementation is available as either Javascript library or Silverlight 3.0 component. We have chosen the Javascript version to eliminate the need for end users to install Silverlight in order to view the images. Zoom-in loads higher resolution image tiles saved as separate JPEG files. This provides a smooth transition when switching between different levels of resolution (Figure 2).

### Results

Although Snapshot Creator and NDPI-Splitter share common libraries to handle the manipulation of proprietary NDPI files, they are used in different contexts.

### Snapshot Creator

Snapshot Creator is a Windows batch file and performs a number of operations in a set order, and the resultant image files are moved into designated folders. The Nano-Zoomer scanner is programmed to save the images in the 'NDPI-New' folder (Figure 3), from where they are processed overnight and are moved to 'NDPI-Processed' folder. If any errors occur, the images are moved to the 'NDPI-Failed' folder. The scanned images are given the same name as their respective slide identifiers in the ABCTB database. The middle section of the image is captured from the processed NDPI file as a JPEG snapshot, at a magnification level defined in the configuration file ('snapshot-creator.properties') of the software. The ABCTB logo is added as a watermark on the bottom of the JPEG snapshot. Snapshot Creator then links the image to the respective specimen in the Caisis database. Once the JPEG snapshot has been successfully linked to the database, the images are moved into the 'JPEG-Processed' folder. Successfully linked images are run through the Deep Zoom converter tool, which creates tiles of the image at different resolutions. These are published to a directory visible to the website. If linking to the specimen fails, the images are moved to the 'JPEG-Failed' folder and an automatic email is sent to the database administrator. The failed images are looked at manually and if failure is due to a incorrect filename then after correcting the file names, images are put back into the 'JPEG Snapshot Processing' to be processed the next night. All folder paths, database locations, magnifications and JPEG quality are configurable through the snapshot-creator.properties file. The process is diagrammatically explained in Figure 3.

### Integration with Caisis

After the successful creation of JPEG images from the NDPI files, Snapshot Creator links the snapshots to the database using the filename of the image. The filename is an identifier of the slide in the ABCTB customised open source cancer-research database, Caisis [7], wherein histology slides are recorded as specimens. A stored procedure updates the matched field of the Specimens table. Implementation of Snapshot Creator in other SQL databases can be achieved, as this is configurable in the configuration file (snapshot-creator.properties), however, the SQL stored procedures have to be re-written to match the database structure of the prospective system. A full customized copy of Caisis, in use by the ABCTB, can be requested from the corresponding author. Source code of Caisis and our tools is available under GNU General Public Licence [8].

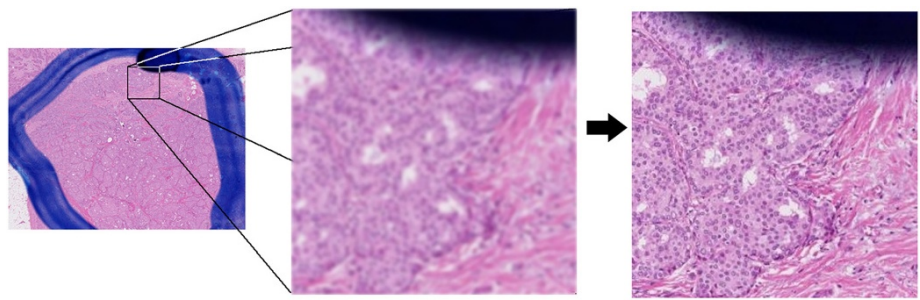

**Figure 2 Deep Zoom overview: Using Deep Zoom the users can drill down to a higher level of zoom without any progress bar wait.** The high resolution image loads in the background and seamlessly blends over the lower resolution image.

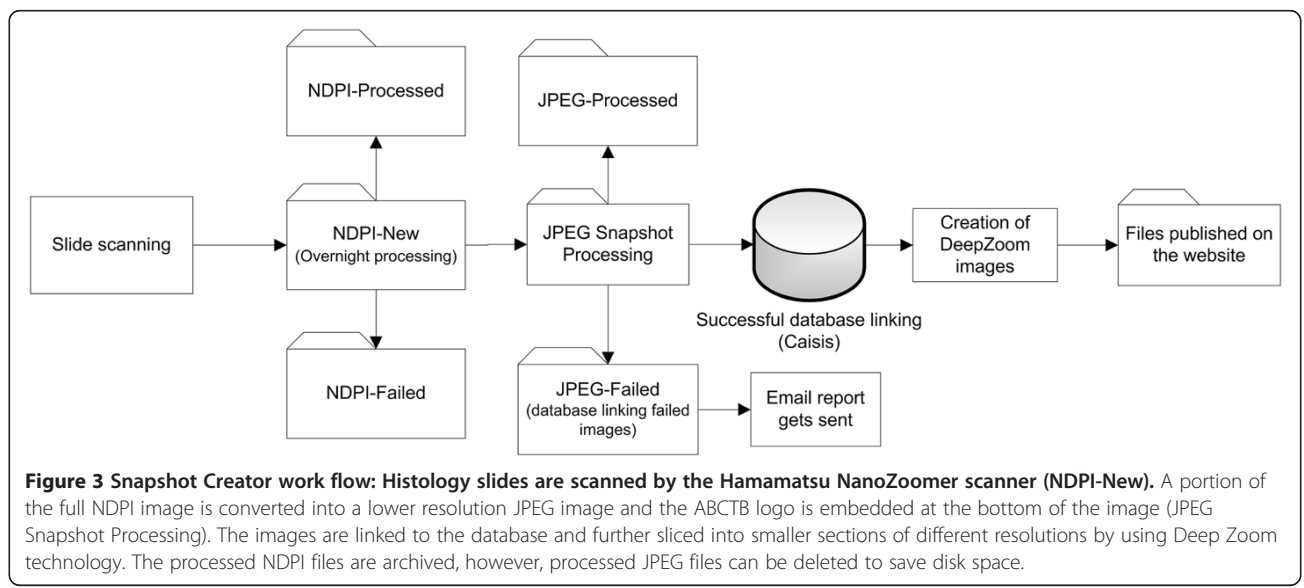

**Figure 3 Snapshot Creator work flow: Histology slides are scanned by the Hamamatsu NanoZoomer scanner (NDPI-New).** A portion of the full NDPI image is converted into a lower resolution JPEG image and the ABCTB logo is embedded at the bottom of the image (JPEG Snapshot Processing). The images are linked to the database and further sliced into smaller sections of different resolutions by using Deep Zoom technology. The processed NDPI files are archived, however, processed JPEG files can be deleted to save disk space.

### NDPI-Splitter

Like Snapshot Creator, NDPI-Splitter is a Windows application, and it is designed to split NDPI image files into smaller TIFF tiles, which can subsequently be imported into image analysis software for automatic processing. In "Step 1" a graphical user interface (GUI) prompts the user to select one or multiple files (Figure 4). Once the files are selected the GUI displays the size and magnification of the image. On the "Step 2" screen (Figure 5) the user can choose the desired width and height of the sections in pixels and magnification level. There is also an option to have empty images filtered out, based on two algorithms i) intensity based: this algorithm is best suited to black/fluorescent images, ii) compression based: this is best suited to images with white backgrounds. These algorithms are designed specifically for the type of images produced by the NanoZoomer. The intensity algorithm examines the intensity of each pixel in an image section, to determine whether each pixel is close to black or not. Based on an average threshold and whiteness threshold, the algorithm determines whether the section contains sufficient pixels of adequate intensity to be interesting. The compression based algorithm uses JPEG compression to compress the image section. If a high level of compression is achieved, the algorithm extrapolates that most of the image is white and therefore is not of interest. The thresholds used by these algorithms can be customised to fine-tune the results. For each image a directory is created containing the TIFF tiles using a grid-based position-dependent naming convention for the newly generated TIFF files. For example the tiles from the first logical row division of the image are named by appending letters such as A1, A2, A3 and so on.

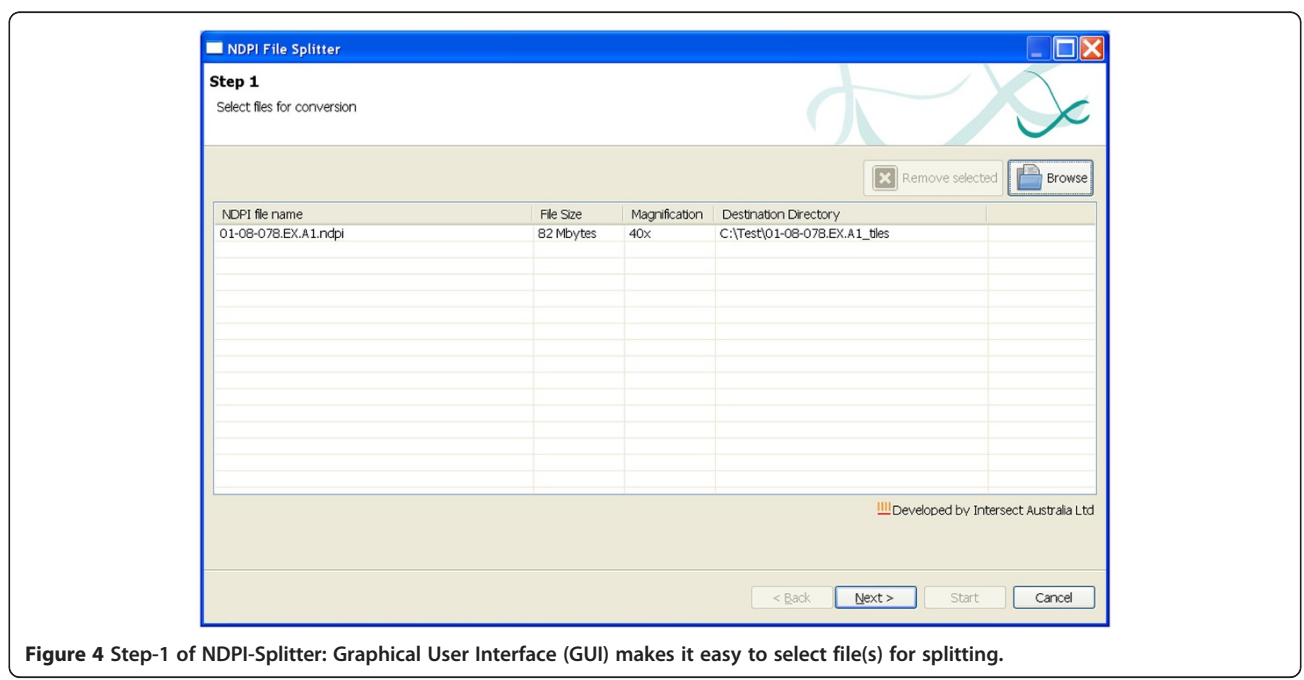

**Figure 4 Step-1 of NDPI-Splitter: Graphical User Interface (GUI) makes it easy to select file(s) for splitting.**

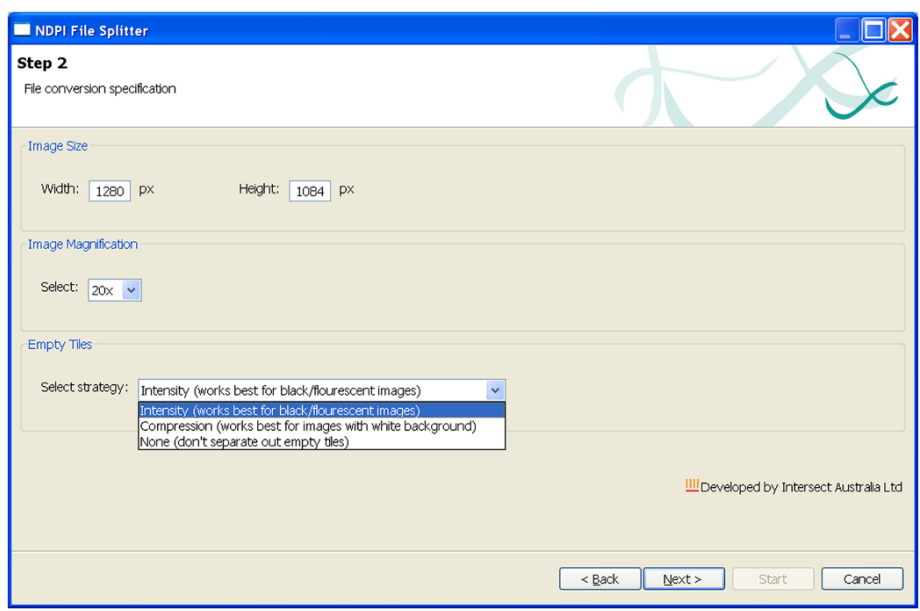

**Figure 5** Step-2 of NDPI-Splitter: At step 2 the width and height of each image section, and the magnification level, can be defined, and there is an option to filter out empty images.

If an empty tile algorithm is selected, the tiles that are determined by the algorithm to contain no digital pathology information are placed in a sub-directory called "empty_tiles". A log file called "log.txt" is created which records the calculation used to decide if a tile is empty or not. The log file can be used to fine-tune the thresholds for assigning 'empty' status. The "emptiness" algorithms, which are based on a determination of the numbers of pixels diverging from a threshold background value, are not perfect and the accuracy of the results varies for different images. For this reason the 'empty' tiles are retained for review, in case tiles that are not empty have been discarded. The application's default values can be configured through a properties file named NDPIsplitter.properties.

## Discussion

Snapshot Creator and NDPI-Splitter are developed in Java and share common libraries to interact with the NDPI files. Technically they are very similar; however they are used in two very different contexts. Snapshot Creator is used to publish lower resolutions JPEG images on a tissue bank web search engine. On the other hand, NDPI-Splitter produces files that can be imported into sophisticated image analysis packages such as Metamorph. The current limitation of image analysis software is their dependency on computer specifications, and typically large images fail to be processed because of insufficient memory. In addition, the significant time required for extraction of TIFF images from large image files is a significant limitation in image analysis. Therefore the ability of NDPI-Splitter to split large files into smaller TIFF sections enables their import into and analysis by image analysis software.

Previous studies have reported different ways of automatic image analysis on virtual slides by identifying regions of interest [9-11]. For example, Romo et al. [10] employed colour, intensity, orientation and texture to calculate a relevance score against a manually selected region of interest. However, by contrast, NDPI-Splitter does not identify regions of interest, rather it creates files that can be imported into automated image analysis pipelines. In addition, NDPI-Splitter, using intensity- and compression-based algorithms, can identify 'empty' regions that contain no or few pixels, which is a novel feature that streamlines the process of importing files for image analysis. This strategy reduces the requirement for manual review of tiles prior to image analysis and minimizes the input to downstream analysis, representing a significant time saving.

Snapshot Creator produces a snapshot, representing one quarter of the full slide image, for publishing on the website, allowing researchers to use the online image search engine and image viewer to determine rapidly whether the biobank holds the material they are interested in. If researchers are interested in applying to the bank for full scanned slides and related datasets, they can do so based on their rapid search of the online images, and full applications are assessed by peer-review [12].

Snapshot Creator takes the snapshot from the middle of the slide, in order to maximise the chance of including the cancer/malignant region in the snapshot, and in approximately 95% of our published images the malignant section is indeed present. However, for slides where the cancer region is markedly offset on the slide, the cancer region can be missed. In order to avoid this, all newly published images are manually reviewed. If the malignant region is not in the snapshot, a manual snapshot of the image is taken, and placed into the 'JPEG Snapshot Processing' folder (Figure 3) for processing the next night.

We have used Deep Zoom for publishing images on the website. Although there are a number of other options, such as the server-side software Spectrum (Aperio Inc.), SlidePath (Digital Pathology Solutions, Ireland) or NDP. Serve (Hamamatsu, Japan), these server-side proprietary products are very expensive. In addition, keeping full-sized images on a storage location accessible from a webserver may not be desirable due to the expensive nature of such storage systems. As an alternative to Deep Zoom, Lien et al. [13] developed a web-based solution for viewing large sized microscopic images derived from the Aperio ScanScope, however, the availability of such code, and maintenance of its currency, could be a limitation to its widespread use. An open source solution such as Deep Zoom from Microsoft is more widely accessible, and it is more likely that it will be maintained in the future. Deep Zoom provides the options of implementation either via Silverlight or JavaScript. We have used the JavaScript version because all modern browsers have built-in JavaScript support, whereas most web browsers do not have the Silverlight plugin installed. Zoomify EZ is another alternative to Deep Zoom, however this requires Flash plugin. Zoomify and Deep Zoom use the same underlying algorithm where images are cut into small sections of images for different resolutions and saved into a logical folder structure.

Caisis is an open source cancer research database, with built-in fields for various cancers such as adrenal, bladder, colon, kidney, penile, prostate, testicular, breast, urological, pancreas and bladder, and the addition of more diseases or new fields is easily achievable. We have customised Caisis to link snapshots and virtual slides derived from the Snapshot Creator and NDPI-Splitter tools. Using Caisis, images can be searched for based on patient history, treatment or biomarkers, and relevant images can then be easily identified and sent to researchers. Therefore Snapshot Creator, NDPI-Splitter coupled with Deep Zoom and the customised Caisis database provide complete management of virtual images. Other researchers have also indicated the future development of such tools [14], therefore our open source tools provide the research community an alternate solution to in-house development.

In summary, as virtual microscopy is moving into the main stream of diagnostic pathology, teaching and research [15], the development of open source tools that manage, catalogue and process virtual slides are needed. A web search engine holding digitized images can be used in teaching environments, to illustrate normal and abnormal cell structures of different cancer type, such as invasive or *in situ* cancer, and is broadly available for research and clinical pathology review. Therefore, NDPI Splitter, Snapshot Creator, Caisis and Deep Zoom are open source tools that provide the ability to make greater use of digital images and therefore broaden the range of applications for tissue bank images.

### Availability and requirements

- Project name: NDPI-Splitter
- Project home page: http://code.google.com/p/NDPI-Splitter/, a full customised copy of Caisis, in use by the ABCTB, can be requested from the corresponding author.
- Operating system(s): Windows
- Programming language: Java
- Requirements: Java, Hamamatsu SDK, JAI 1.1.3, JAI Image IO 1.1, Ant, Deep Zoom
- License: GNU GPL version 3 [8]

### Ethical approval

Australian Breast Cancer Tissue Bank (ABCTB) is covered by Protocol No: X12-0279 Ethics Review Committee, Royal Prince Alfred Hospital, Camperdown, NSW 2050 Australia.

**Author details**
[1]Australian Breast Cancer Tissue Bank, University of Sydney at the Westmead Millennium Institute, Darcy Road, Westmead, NSW 2145, Australia. [2]Intersect

Australia, Sydney, NSW, Australia. ³Westmead Institute for Cancer Research, Sydney Medical School – Westmead, University of Sydney at the Westmead Millennium Institute, Westmead, NSW 2145, Australia.